\documentclass[12pt]{article}
\usepackage{authblk}
\usepackage[T1]{fontenc}
\usepackage[utf8]{inputenc}
\usepackage[margin=0.75in]{geometry}
\usepackage[square,numbers]{natbib}

\title{\Large \textbf{Strain fluctuations unlock ferroelectricity in wurtzites}}

\author[1]{Steven M. Baksa}
\author[1]{Simon Gelin}
\author[1]{Seda Oturak}
\author[1]{R. Jackson Spurling}
\author[2]{Alireza Sepehrinezhad}
\author[1]{Leonard Jacques}
\author[1]{Susan E. Trolier-McKinstry}
\author[3]{Adri C. T. van Duin}
\author[1]{Jon-Paul Maria}
\author[4]{Andrew M. Rappe}
\author[5]{Ismaila Dabo}

\affil[1]{Department of Materials Science and Engineering, Pennsylvania State University, University Park, Pennsylvania, United States, 16802}
\affil[2]{Department of Engineering Science and Mechanics, Pennsylvania State University, University Park, Pennsylvania, United States, 16802}
\affil[3]{Department of Mechanical Engineering, Pennsylvania State University, University Park, Pennsylvania, United States, 16802}
\affil[4]{Department of Chemistry, University of Pennsylvania, Philadelphia, Pennsylvania, United States, 19104}
\affil[5]{Department of Materials Science and Engineering, Carnegie Mellon University, Pittsburgh, Pennsylvania, United States, 15213}
\date{}

\usepackage{gensymb}
\usepackage{braket}
\usepackage{amsmath,amsfonts}
\usepackage{graphicx}
\usepackage{makecell}
\usepackage{float}
\usepackage[colorlinks=true, allcolors=blue]{hyperref}
\usepackage{caption,subcaption}

\renewenvironment{abstract}
 {\small
  \begin{center}
  \bfseries \abstractname\vspace{-.5em}\vspace{0pt}
  \end{center}
  \list{}{
    \setlength{\leftmargin}{.20in}%
    \setlength{\rightmargin}{\leftmargin}%
  }%
  \item\relax}
 {\endlist}

\DeclareUnicodeCharacter{2009}{\,}

\begin{document}

\maketitle

\begin{abstract}
    \noindent Ferroelectrics are of practical interest for non-volatile data storage due to their reorientable, crystallographically defined polarization. Yet efforts to integrate conventional ferroelectrics into ultrathin memories have been frustrated by film-thickness limitations, which impede polarization reversal under low applied voltage. Wurtzite materials, including magnesium-substituted zinc oxide (Zn,Mg)O, have been shown to exhibit scalable ferroelectricity as thin films. In this work, we explain the origins of ferroelectricity in (Zn,Mg)O, showing that large strain fluctuations emerge locally in (Zn,Mg)O and can reduce local barriers to ferroelectric switching by more than 40\%. We provide concurrent experimental and computational evidence of these effects by demonstrating polarization switching in ZnO/(Zn,Mg)O/ZnO heterostructures featuring built-in interfacial strain gradients. These results open up an avenue to develop scalable ferroelectrics by controlling strain fluctuations atomistically. 
\end{abstract}

\section{Introduction}

Ferroelectric thin films can reverse their crystallographically defined polarization states under moderate electric bias, making them attractive candidates for storing binary data in energy-efficient (low-voltage) microelectronics \cite{Dawber2005, Mikolajick2020}. Cation substitution is an effective approach for promoting polarization reversal in prototypical ferroelectrics like titanate-based perovskite oxides \cite{Cohen1992, Xu2022}. For instance, Pb(Zr,Ti)O$_3$ (Zr-substituted PbTiO$_3$) exhibits enhanced polarizability relative to its PbTiO$_3$ and PbZrO$_3$ end members~\cite{Izyumskaya2007} due to competing polar structures at the morphotropic (tetragonal-to-rhombohedral) phase boundary. This polar-phase instability leads to a high dielectric permittivity, making it an appealing ferroelectric material \cite{Nuraje2013}.

\begin{figure}
    \centering
    \includegraphics[scale=0.80]{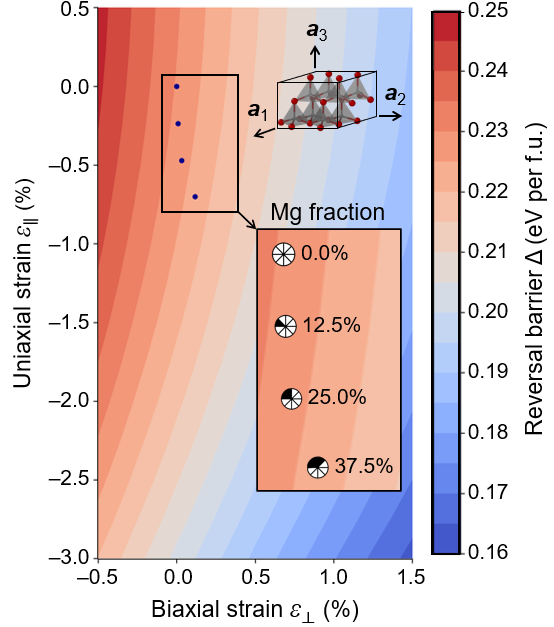}
    \caption{Polarization reversal barriers as a function of biaxial tension $\varepsilon_\perp$ and uniaxial compression $\varepsilon_\parallel$ (orthogonal and parallel to the polar axis, respectively; as shown in the inset) calculated from first principles. Lattice strains $\varepsilon_\perp$ and $\varepsilon_\parallel$ associated with the lowest-energy, fully relaxed (Zn,Mg)O geometries relative to bulk ZnO up to the solubility limit of Mg (ca.~40\%) are also shown as points.}
    \label{fig:Strained-ZnO}
\end{figure}

While Pb(Zr,Ti)O$_3$ has been successfully integrated into random-access ferroelectric memories \cite{Morrison2007}, there exist scaling constraints for film thicknesses below 70 nm, which limit the ferroelectric performance and energy efficiency of Pb(Zr,Ti)O$_3$-based thin-film microelectronics \cite{Ma2002, Kohlstedt2005, Kim2005, Nagarajan2005, Ihlefeld2016}. As an alternative to perovskite materials, fluorites \cite{Park2020, Park2022, Kim2023}, including hafnia (HfO$_2$) and its derivatives (for example, (Hf,La)O$_2$, (Hf,Zr)O$_2$, and (Hf,Gd)O$_2$) \cite{Park2018, Song2021, Schroeder2018, Mueller2012}, have been shown to be scalable down to 2 nm \cite{Song2021, Schroeder2018, Mueller2012, Mohan2022, Jung2022, Zhou2022, Lomenzo2020, Skopin2022, Cheema2022}. Although endurance of up to 10$^{10}$ cycles have been reported \cite{Park2018, Song2021, Jung2022}, ferroelectricity in hafnia is typically optimized at synthesis temperatures above 800 $\degree$C to reduce wake up from a pristine to cycled state, whereas back-end-of-the-line processing necessitates deposition temperatures below 400 $\degree$C. Studies into (Hf,Zr)O$_2$ thin films at low annealing temperatures (<400 $\degree$C) show ferroelectric responses with a remanent polarization of 23.5 $\mu$C$\cdot$cm$^{-2}$ and coercive voltage of 1.25 V \cite{Mohan2022, Yu2022}.

Unlike perovskites and fluorites, wurtzites \cite{Kim2023, Konishi2016, Moriwake2014, Moriwake2020, Calderon2023} can be synthesized under nominal processing conditions and exhibit a stable polar phase with no observable Curie temperature below their melting point \cite{Shimizu2016}. Although wurtzites tend to have significant polarization reversal barriers, a dopant atom can stress the polar structure to the point of inducing a ferroelectric response at low bias, as is the case for (Al,Sc)N, which exhibits a remanent polarization of 80 $\mu$C$\cdot$cm$^{-2}$ with a coercive field of 3.1 MV$\cdot$cm$^{-1}$ \cite{Yazawa2021, Yazawa2022, Drury2021, Wolff2021}. Similarly, while ZnO adopts a polar wurtzite structure that is generally not considered to be ferroelectric due to its high coercive field (>7 MV$\cdot$cm$^{-1}$) \cite{Konishi2016}, it has been shown experimentally that (Zn,Mg)O is ferroelectric with a remanent polarization over 100 $\mu$C$\cdot$cm$^{-2}$ (80 $\mu$C$\cdot$cm$^{-2}$ from first principles) and a coercive field as low as 2.7 MV$\cdot$cm$^{-1}$ \cite{Ferri2021} after less than forty electric field cycles \cite{Jacques2023}, although the underlying mechanisms of this induced ferroelectricity remain elusive. 

Although electrostatic effects due to domain wall propagation in the thin film have been suggested as a possible mechanism, it has been proposed that epitaxial strain can impart ferroelectricity to polar materials like wurtzite ZnO (Figure \ref{fig:Strained-ZnO}) \cite{Konishi2016, Moriwake2014, Moriwake2020, Yazawa2021}. For instance, a biaxial tension orthogonal to the polar axis of 1.5$\%$ can reduce the reversal barrier by 20$\%$. First-principles studies have suggested that a uniaxial compression of 6.0 GPa along the [0001] direction stabilizes an intermediate hexagonal phase in ZnO \cite{Sarasamak2008}. Using this insight, strain induced by Mg dopants has been considered as a possible mechanism for ferroelectric reversal in (Zn,Mg)O; however, the magnitude of this biaxial strain is below 0.3\%, which cannot explain such a large change in the ferroelectric switching barriers. 

In this work, this mechanistic discrepancy is resolved by showing that local strain fluctuations can unlock ferroelectric switching in (Zn,Mg)O. The possibility to promote ferroelectricity \textit{via} local-strain control offers a versatile strategy to develop and process low-temperature scalable ferroelectric thin films and heterostructures for microelectronic memories and switches. 

\section{Results and discussion}
\subsection{Influence of cation substitution on strain fluctuations}

\begin{figure}
    \centering
    \includegraphics[scale=0.80]{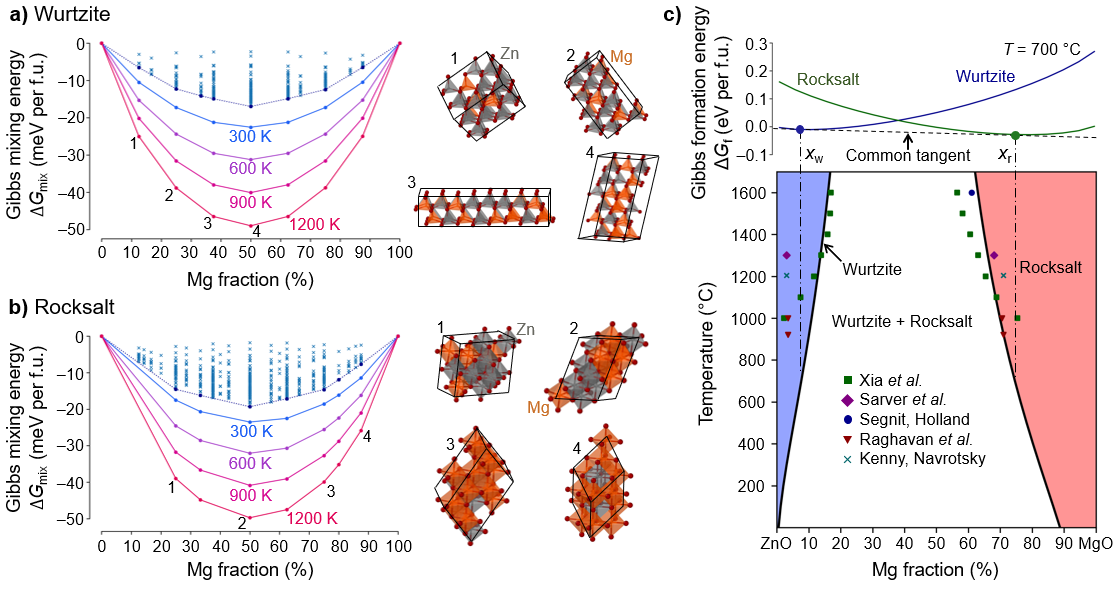}
    \caption{Convex hulls and representative structures at finite temperature of the (a) wurtzite and (b) rocksalt phases computed from density-functional theory. (c) Calculated phase diagram of the MgO binary constructed from cluster expansions and Monte Carlo simulations {\it via} Wang-Landau sampling. The Gibbs free formation energies, $\Delta G_{\rm f}$, of each phase were employed to calculate the phase boundaries, $x_{\rm w}$ and $x_{\rm r}$, as a function of temperature. Computed phase boundaries and experimental data (Refs.~\cite{Xia2016, Sarver1959, Segnit1965, Raghavan1991, Kenny1972}) are indicated with solid lines and points, respectively.}
    \label{fig:phase-diagram}
\end{figure}

The formation of (Zn,Mg)O is critically dependent on the solubility of Mg cations in ZnO. To predict this solubility, the mixing enthalpy of (Zn,Mg)O as a function of Mg content for the end-member (wurtzite and rocksalt) phases was calculated as $\Delta H_{\rm mix} (x) = H_{{\rm Zn}_{1-x}{\rm Mg}_x{\rm O}} - (1-x) H_{\rm ZnO} - x H_{\rm MgO}$, where $x$ is the Mg molar fraction, $H_{{\rm Zn}_{1-x}{\rm Mg}_x{\rm O}}$ is the energy of the (Zn,Mg)O configuration per formula unit, and $H_{\rm ZnO}$ and $H_{\rm MgO}$ are the reference enthalpies of ZnO and MgO, respectively. The results of these calculations, shown in Figure \ref{fig:phase-diagram}(a) and (b), indicate that mixing is favored across the full composition range for the wurtzite and rocksalt structures. Yet the shallow depths of --17.5 and --20 meV per formula unit (comparable to the thermal energy under processing conditions) for the wurtzite and rocksalt phases, respectively, and the existence of multiple configurations near the convex hull suggest that thermal disorder may impact local structural motifs in (Zn,Mg)O.

To include thermal effects, a cluster expansion model \cite{ICET, Sanchez1984, ATAT} was constructed and Wang--Landau sampling \cite{Wang-Landau1, Wang-Landau2, Landau2004} was applied to predict the temperature-dependent Gibbs free energies of (Zn,Mg)O configurations containing thousands of atoms. (The description of the cluster expansion model and subsequent Monte Carlo simulations is provided in Section S2.2 of the supplementary information, SI.) This stochastic sampling approach simulates the microcanonical density of states as a function of the ensemble energy and provides accurate estimates for the configurational entropy assuming vibrational entropy to be negligible. The resulting free energies, along with representative structures, are shown in Figure \ref{fig:phase-diagram}.

By drawing common tangents between the free energies of formation as a function of Mg molar fraction at different temperatures [Figures \ref{fig:phase-diagram}(a) and \ref{fig:phase-diagram}(b)], a phase diagram was constructed for the MgO mixture [Figure \ref{fig:phase-diagram}(c)]. The predicted phase diagram is in close agreement with experimental data \cite{Xia2016, Sarver1959, Segnit1965,Raghavan1991,Kenny1972} and compares favorably with previous first-principles predictions \cite{Liu2012, Koster2015} using a single adjustable parameter (a uniform shift in the formation energies of the rocksalt configurations by 0.19 eV per formula unit, f.u.). On average, errors on relative formation energies are in the range of 0.2-0.8 eV per atom with a mean absolute deviation of 0.26 eV per atom due to incomplete error cancellation of energy differences within density functional theory under the generalized gradient approximation \cite{Stevanovic2012}. This single-parameter correction is thus consistent with the typical error margins of density-functional-theory calculations.

\begin{figure}
    \centering
    \includegraphics[scale=0.85]{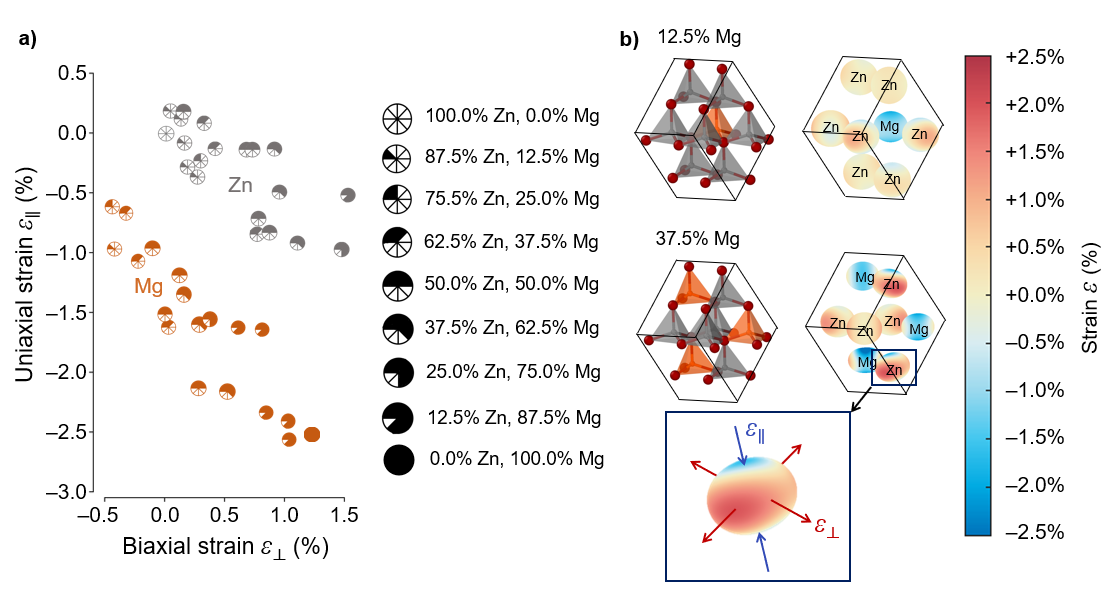}
    \caption{(a) Calculated atom-resolved strains around  Zn (grey) and Mg (orange) cations as a function of Mg fraction. The Zn cations have a mixture of $\varepsilon_\perp$ and $\varepsilon_\parallel$, whereas Mg cations predominantly have $\varepsilon_\parallel$. The strain magnitude increases with Mg fraction and significant local strain fluctuations occur. (b) Illustration of (Zn,Mg)O structures at 12.5$\%$ and  37.5$\%$ Mg fraction in addition to the corresponding strain ellipsoids centered around their respective cations. The inset includes an illustration of a strain ellipsoid, which represents the rigid rotation and elastic deformation of the local bonding environment around individual Zn and Mg cations at the atomic scale.}
    \label{fig:local_strain_analysis}
\end{figure}

To quantify Mg-induced distortions, the local strain around individual Zn and Mg cations was calculated by examining the deformation of the coordination environment of the $N$ nearest O neighbors ($N=4$), which are initially located at positions $\boldsymbol{R}_n$ (in the ideal ZnO geometry) and are then displaced to the new positions $\boldsymbol{r}_n$ (in the distorted (Zn,Mg)O geometry):
\begin{equation}
    \boldsymbol{\varepsilon} = \frac{1}{2} \left(\boldsymbol{J} \boldsymbol{J}^{\top} - \boldsymbol{I} \right) = \frac{3}{2\bar r^2 }\langle \boldsymbol r \boldsymbol r^\top \rangle- \frac{\boldsymbol{I}}{2},
    \label{eqn:local_strain}
\end{equation}
where $\langle \boldsymbol r \boldsymbol r^\top \rangle$ stands for $\frac 1N {\sum_{n=1}^N \boldsymbol{r}_n \boldsymbol{r}_n^{\top}}$, $\boldsymbol{J} = {\rm arg \, min}_{\boldsymbol{j}} \sum_{n=1}^N (\boldsymbol{r}_n - \boldsymbol{j} \boldsymbol{R}_n)^2 $ is the (best-fit) deformation matrix from $\boldsymbol{R}_n$ to $\boldsymbol{r}_n$, $\boldsymbol{I}$ denotes the identity matrix, and $\bar r$ is the bond length in the idealized geometry, where all undistorted bonds have the same length $\bar r = \boldsymbol{R}_n $. (The derivation of Eq.~\ref{eqn:local_strain} is presented in Section S2.3 of the SI.)

Graphically, these local deformations can be visualized as colored ellipsoids, whose shape represents the directional compression (in blue) or expansion (in red) of the Zn and Mg coordination environments in three dimensions, as shown in Figure \ref{fig:local_strain_analysis}. Notably, Figure \ref{fig:local_strain_analysis}(a) reveals that Zn and Mg cations have distinct strain patterns. The strain surrounding Zn cations is a mixture of $\varepsilon_\perp$ and $\varepsilon_\parallel$, whereas that around Mg cations is predominantly of $\varepsilon_\parallel$ character. This phenomenon has previously been reported in similar cations in perovskite oxides \cite{Grinberg2005}. As the Mg fraction increases, the magnitudes of $\varepsilon_\perp$ and $\varepsilon_\parallel$ increase, showing that Zn and Mg experience higher local strain in (Zn,Mg)O. Fluctuations of $\varepsilon_\perp$ and $\varepsilon_\parallel$ at intermediate concentrations are significantly larger than at dilute concentrations. These fluctuations can reach up to 1.5$\%$ in $\varepsilon_\perp$ and --2.5$\%$ in $\varepsilon_\parallel$, while the global strain (that is, their macroscopic average across the periodic supercell) never exceeds 1.0$\%$. Visualizing the strain ellipsoids [Figure \ref{fig:local_strain_analysis}(b)] confirms that Mg cations exhibit significantly stronger uniaxial compression than Zn cations, whereas Zn cations exhibit marginally stronger biaxial tension than Mg cations. (A Wannier-function analysis of the chemical bonding in (Zn,Mg)O is presented in Section S2.4 of the SI.) In the next section, we examine how the local strains, which are highlighted in Figure \ref{fig:local_strain_analysis}, ultimately affect polarization switching in (Zn,Mg)O.

\subsection{Influence of strain fluctuations on ferroelectricity}
\label{sec:strain-ferroelectricity}
\begin{figure}
    \centering
    \includegraphics[scale=0.90]{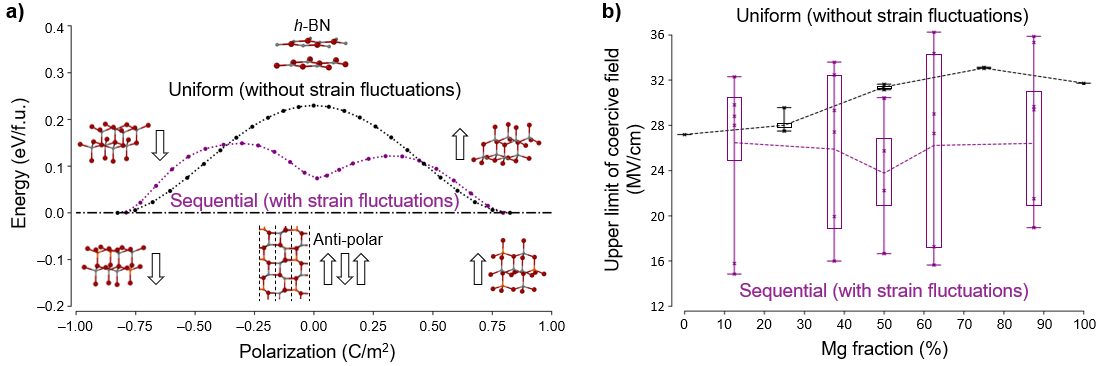}
    \caption{(a) Minimum energy pathways of (Zn,Mg)O simulated using nudged-elastic-band calculations. Here, Mg fractions of 0\% and 37.5\% are shown to illustrate the effect of strain fluctuations on the ferroelectric polarization reversal. While uniform reversal (black) passes through a hexagonal boron nitride ($h$-BN) intermediate structure, sequential reversal (purple) passes through an anti-polar structure with alternating polar states. Sequential reversal reduces the energy barrier between the bistable states by 40\%. (b) Estimates of the coercive field and its variance under uniform and sequential reversal. If a structure contains larger strain fluctuations, there is a strong likelihood that it will undergo a sequential ferroelectric reversal under reduced coercive fields. Unless the minimum energy pathway is the actual pathway, these calculations of the coercive fields are an upper limit of the experimental coercive field.}
    \label{fig:NEB_selected_results}
 \end{figure}

 To determine the minimum energy pathways of ferroelectric polarization reversal in (Zn,Mg)O, we performed nudged-elastic-band calculations \cite{Jonsson1998, Henkelman2000, Henkelman2000_2} at both 0\% Mg and 37.5\% Mg doping [Figure \ref{fig:NEB_selected_results}(a)]. In the case of 0\% Mg (pure ZnO, without local strain fluctuations), we observed uniform switching where all cations move simultaneously in one step. At 37.5\% Mg, the structure engages in sequential switching where cations move consecutively in multiple steps due to differences in their coordination environment (the local strain fluctuations). This type of sequential reversal mechanisms is characterized by intermediate states passing through a non-polar $h$-BN crystal structure for uniform reversal and an anti-polar crystal structure with alternating polar states for sequential reversal, as recently observed experimentally and computationally in a related wurtzite material, (Al,B)N \cite{Calderon2023}. 

 In quantitative terms, sequential reversal arising from local strain fluctuations leads to a 40$\%$ reduction in the energy barrier compared to uniform reversal. This reduction in the reversal barrier is further validated when considering the local coercive fields, which were calculated by applying a multi-Gaussian fitting on the nudged-elastic-band pathways. As reported in Figure \ref{fig:NEB_selected_results}(b), uniform reversal with little strain fluctuations leads to high coercive fields with little variance, whereas sequential reversal with significant fluctuations is characterized not just by lower coercive fields, but also by a larger variance of these fields, with a reduction of up to 45\%. With these large differences in local coercive fields, there are opportunities for the most favorably distorted cation to switch first, prompting nearby cations to switch consecutively. 

\begin{figure}
    \centering
    \includegraphics[width=\textwidth]{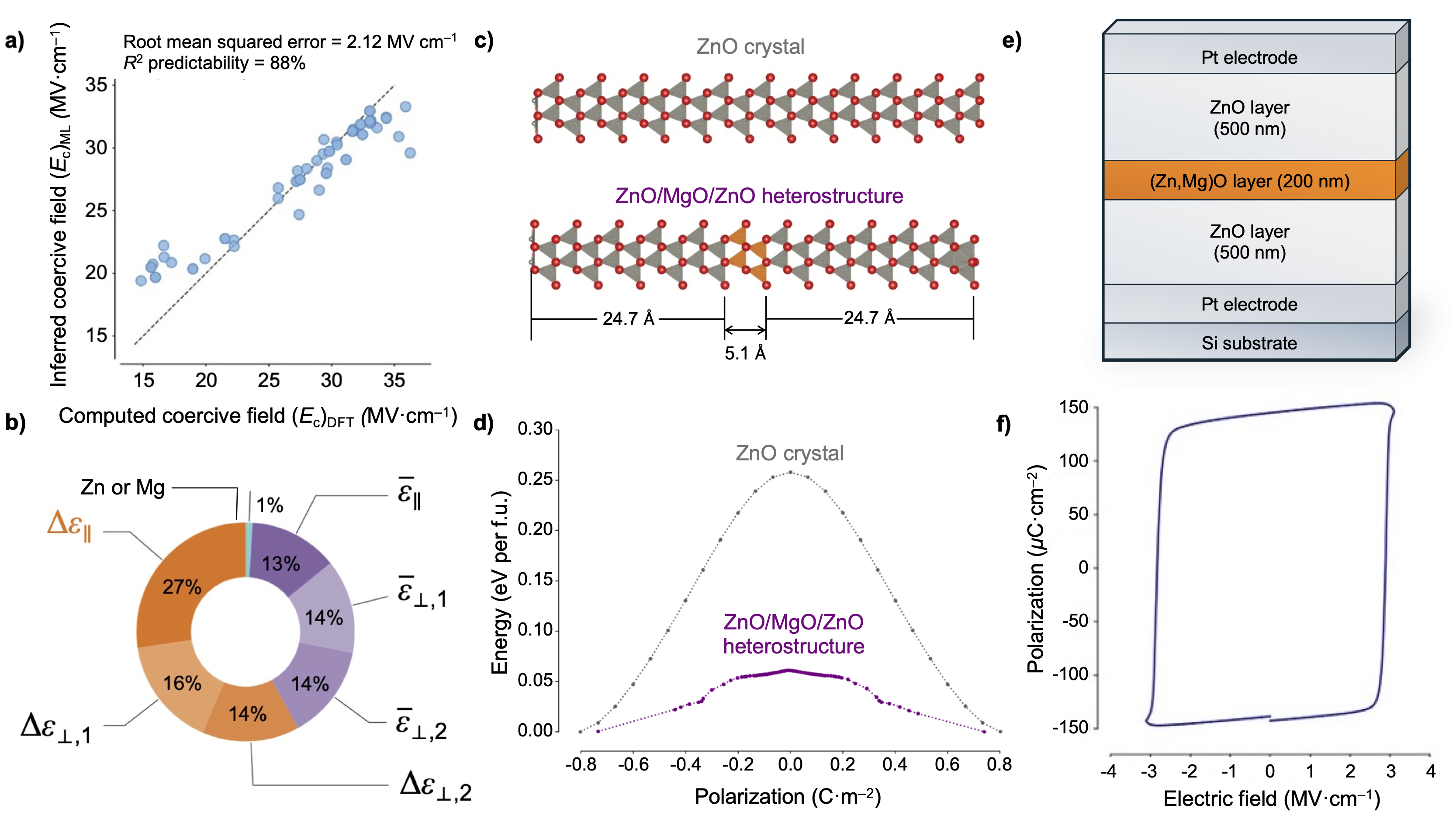}
    \caption{Analysis of local coercive fields and ferroelectric switching in ZnO-based heterostructures. (a) Comparison of the computed and predicted reversal barriers from the multivariate regression model. The regression model predicts local coercive fields with a root mean square error of 2.12 MV cm$^{-1}$ and a coefficient of determination of 88\%. 
    (b) Feature importance ranking of the regression model including changes in strain between the bistable states ($\Delta\varepsilon$) (in orange) and mean strains ($\bar{\varepsilon}$) (in purple). The most important features are changes in the uniaxial ($\Delta \varepsilon_{\parallel}$) and biaxial strain ($\Delta \varepsilon_{\perp}$). The least important feature of the model was the atom type ({\it e.g.}, Zn or Mg). (c) Atomistic model of a MgO:ZnO heterostructure where a layer of MgO (5.1 \AA) is sandwiched between two layers of ZnO (24.7 \AA). (d) Minimum energy pathways of (Zn,Mg)O simulated using machine-learning interatomic potentials; while pristine ZnO engages in a uniform reversal, the heterostructure switches \textit{via} domain wall motion, highlighting the effect of length scales in accurately modeling reversal pathways. (e) Schematic of (Zn,Mg)O:ZnO heterostructure film stack with Pt electrodes and Si substrate. (f) Polarization-field measurements showing ferroelectric hysteresis for the heterostructure film stack.}
    \label{fig:ZMO-panel}
\end{figure}

To confirm the predominance of strain fluctuations in inducing ferroelectricity in (Zn,Mg)O, a simple multivariate regression (a random forest model) \cite{breiman2001random, molnar2020interpretable} was applied to the local coercive fields calculated in Figure \ref{fig:NEB_selected_results}. In this model, the local strains in the negatively and positively poled states of each cation were parsed and combined into nine input features, namely, the strain in the initial and final poled states ($\varepsilon^+_\parallel$, $\varepsilon^+_\perp$, $\varepsilon^-_\parallel$, and $\varepsilon^-_\perp$), the changes in strain between the bistable states ($\Delta \varepsilon_\parallel = \varepsilon^+_\parallel - \varepsilon^-_\parallel$ and $\Delta \varepsilon_\perp = \varepsilon^+_\perp - \varepsilon^-_\perp$), the mean strains [$\overline{\varepsilon}_\parallel = (\varepsilon^+_\parallel + \varepsilon^-_\parallel)/2$ and $\overline{\varepsilon}_\perp  = (\varepsilon^+_\perp + \varepsilon^-_\perp)/2$], and the identity of the switching cation (Zn or Mg). Training and validation of the regression model are presented in Figure \ref{fig:ZMO-panel}(a), resulting in a root mean square error (RMSE) of 2.12 MV$\cdot$cm$^{-1}$ and a high coefficient of determination ($R^2$) of 88\%. Discrepancies between simulations and the regression model are limited to the lowest and highest coercive fields, suggesting that second-nearest-neighbor strain fluctuations may influence local coercive fields to some extent. 

Feature importance ranking [Figure \ref{fig:ZMO-panel}(b)] reveals that the dominant contributors to the coercive field are the changes in the uniaxial compression $-\Delta \varepsilon_{\parallel}$ and biaxial tension $\Delta \varepsilon_{\perp}$ between the poled states. The regression model also reveals that the identity of the cation (Zn or Mg) has little bearing on the coercive field. These results strongly suggest that maximizing $\Delta \varepsilon_{\parallel}$ (that is, maximizing the strain gradient along the polar direction) would be effective at facilitating polarization reversal. 

The hypothesis that polar strain gradient ($\Delta \varepsilon_{\parallel}$) can promote ferroelectricity was critically tested by examining a model heterostructure where a layer of MgO is sandwiched between two layers of ZnO, thereby maximizing $\Delta \varepsilon_{\parallel}$ at the ZnO/MgO interface, as depicted in Figures \ref{fig:ZMO-panel}(c). Figure \ref{fig:ZMO-panel}(d) compares the minimum energy pathway of this heterostructure to that of the same pristine stucture (where Mg is simply replaced with Zn), both calculated using a pre-trained universal machine-learned interatomic potential (the crystal Hamiltonian graph neural network) \cite{deng_2023_chgnet}. These simulations show that, while pristine ZnO engages in a uniform reversal, the heterostructure switches by domain wall motion, with the domain wall region exhibiting an anti-polar structure. It should be noted that the spontaneous polarization of the heterostructure is reduced by 9 $\mu$C$\cdot$cm$^{-2}$ (11\%), which may be attributed to the smaller remanent polarization of $w$-MgO relative to $w$-ZnO.

An experimental realization of this prediction was achieved by depositing a ZnO/(Zn,Mg)O/ZnO  heterostructure, which consists of a $\sim$200 nm thin film of Mg-rich (Zn,Mg)O sandwiched between two $\sim$500 nm thin films of ZnO on a silicon substrate and Pt electrodes, as depicted schematically in Figure \ref{fig:ZMO-panel}(e). Top electrode deposition \textit{via} shadow mask can result in discrepancies in electrode area, thereby inflating the measured polarization. That said, experimental values are close to expectations based on first-principles predictions. Polarization-field measurements indicate ferroelectric hysteresis is obtained throughout the heterostructure [Figure \ref{fig:ZMO-panel}(f)] in contrast with pristine ZnO which does not switch at electric fields below the breakdown voltage. These experimental observations are consistent with modeling expectations, which strongly suggests that local or interfacial strain gradients facilitate switching with reduced coercive fields in ZnO-based wurtzite materials.

\section{Conclusion}
This work has shown that (Zn,Mg)O exhibits local strains that largely exceed their macroscopic average. The computed MgO phase diagram was found to be consistent with experimental data using a single corrective parameter, enabling the modeling of representative local configurations in (Zn,Mg)O. Using the predicted configurations, we unveiled the atomistic mechanisms of ferroelectric switching in this material, demonstrating that Mg cations disrupt chemical bonding along the tetrahedral network; these bond disruptions induce atomic-scale strain fluctuations, which trigger ferroelectricity {\it via} sequential polarization reversal. At the practical level, we demonstrated that controlling cation ordering to maximize local strain gradients (for example, by synthesizing ZnO/(Zn,Mg)O/ZnO heterostructures or by introducing Zn:Mg composition gradients along the polar axis) can induce sequential polarization reversal or domain wall propagation, both mediated by local anti-polar transition states. This general approach provides a broadly applicable design principle to unlock ferroelectricity in wurtzites and related polar materials. 

\section{Acknowledgements}
\noindent \textbf{Funding:} This material is based upon work supported by the US Department of Energy, Office of Science, Office of Basic Energy Sciences Energy Frontier Research Centers program under Award Number DE-SC0021118 (Center for 3D Ferroelectric Microelectronics). All first-principles calculations and analyses were conducted using the Pennsylvania State University ROAR supercomputer resources of the Institute for Computational and Data Sciences.
\newline
\newline
\noindent \textbf{Author contributions:} 
\textit{conceptualization}--S. M. B., A. M. R., S. E. T.-M., I. D.; 
\textit{methodology}--S. M. B., S. G., S. O., R. J. S., A. S., L. J., S. E. T.-M., A. C. T. V., J.-P. M., I. D.; 
\textit{software}--S. M. B., S. G., S. O., A. S., A. C. T. V., I. D.;
\textit{formal analysis}--S. M. B., S. G., S. O., R. J. S., A. S., L. J., S. E. T.-M., A. C. T. V., J.-P. M., I. D.; 
\textit{data curation}--S. M. B.; 
\textit{drafting}--S. M. B., I. D.; 
\textit{review and editing}--all authors; 
\textit{visualization}--S. M. B., S. O., I. D.; 
\textit{project administration}--I. D.; 
\textit{funding acquisition}--S. E. T.-M., J.-P. M., A. M. R., I. D.
\newline
\newline
\noindent \textbf{Conflicts of interest:} The authors declare no conflicts of interest.
\newline
\newline
\noindent \textbf{Data and materials availability:} The data that support the findings of this study are available from the corresponding author upon reasonable request.

\pagebreak
\bibliographystyle{MSP}
\bibliography{references}

\end{document}